\documentclass[secnumarabic,amssymb,pnas,reprint,superscriptaddress,longbibliography]{revtex4-2}
\usepackage[english]{babel}
\usepackage{graphicx}
\usepackage{amsmath}
\usepackage{amsfonts}
\usepackage[usenames,dvipsnames]{color}
\usepackage{hyperref}
\usepackage{blindtext}
\hypersetup{colorlinks}
\usepackage[utf8]{inputenc}
\usepackage{color}
\usepackage[version=3]{mhchem} 
\usepackage{ifthen}
\hypersetup{linkcolor=black, citecolor=blue, urlcolor=blue}
\makeatletter
\addto\captionsenglish{\renewcommand{\figurename}{Fig.}}
\renewcommand*{\fnum@figure}{{\normalfont\bfseries \figurename~\thefigure}}
\renewcommand*{\@caption@fignum@sep}{\textbf{ }}
\makeatother
\setcitestyle{round}
\usepackage{multirow}
\usepackage{booktabs}

\newcommand{\red}[1]{{\color{red}#1}}
\newcommand{\blue}[1]{{\color{blue}#1}}

\begin{document}
\title{Realisation of the Brazil-nut effect in charged colloids without external driving}
\author{Marjolein N. van der Linden}
\thanks{These two authors contributed equally}
\address{Soft Condensed Matter, Debye Institute for Nanomaterials Science, Princetonplein 1, 3584 CC, Utrecht, The Netherlands}
\email{m.n.vanderlinden@utwente.nl}
\author{Jeffrey C. Everts}
\thanks{These two authors contributed equally}
\address{Institute of Physical Chemistry, Polish Academy of Sciences, Kasprzaka 44/52, PL-01-224 Warsaw, Poland}
\address{Institute of Theoretical Physics, Faculty of Physics, University of Warsaw, Pasteura 5, 02-093 Warsaw, Poland}
\email{jeffrey.everts@fuw.edu.pl}
\author{René van Roij}
\address{Institute for Theoretical Physics, Center for Extreme Matter and Emergent Phenomena,  Utrecht University, Princetonplein 5, 3584 CC Utrecht, The Netherlands}
\email{r.vanroij@uu.nl}
\author{Alfons van Blaaderen}
\email{a.vanblaaderen@uu.nl}
\address{Soft Condensed Matter, Debye Institute for Nanomaterials Science, Princetonplein 1, 3584 CC, Utrecht, The Netherlands}

\date{\today}

\begin{abstract}

Sedimentation is a ubiquitous phenomenon across many fields of science, such as geology, astrophysics, and soft matter. Sometimes, sedimentation leads to unusual phenomena, such as the Brazil-nut effect, where heavier (granular) particles reside on top of lighter particles after shaking. We show experimentally that a Brazil-nut effect can be realised in a binary colloidal system of long-range repulsive charged particles driven purely by Brownian motion and electrostatics without the need for activity. Using theory, we argue that not only the mass-per-charge for the heavier particles needs to be smaller than the mass-per-charge for the lighter particles, but that at high overall density, the system can be trapped in a long-lived metastable state, which prevents the occurrence of the equilibrium Brazil-nut effect. Therefore, we envision that our work provides valuable insights into the physics of strongly interacting systems, such as partially glassy and crystalline structures. Finally, our theory, which quantitatively agrees with the experimental data, predicts that the shapes of sedimentation density profiles of multicomponent charged colloids are greatly altered when the particles are charge regulating with more than two ion species involved. Hence, we hypothesise that sedimentation experiments can aid in revealing the type of ion-adsorption processes that determine the particle charge and possibly the value of the corresponding equilibrium constants.
\end{abstract}
\maketitle

\section*{Significance}
Understanding the spatial ordering of dispersed particles caused by gravitational fields is relevant for applications as diverse as the separation of particles, water purification, the isolation of cells from blood, the shelf-life of paints and inks, and stabilising emulsions. When external energy is provided, unusual density distributions have been observed: heavier particles float on top of lighter particles. We provide a lacking experimental demonstration that this effect can also occur in Brownian colloidal systems, without external energy input, in dispersions of charged particles in a low-polar oil. Additionally, we demonstrate that density-profile shapes in mixtures of charged colloids, even if out of equilibrium, can give valuable microscopic information on the equilibrium constants determining the particle charge.

\section*{Introduction}

The effects of gravity on the spatial distribution of matter, such as atoms, small molecules, proteins, colloids, and granular matter, can lead to a wide variety of phenomena relevant to science and technology. Understanding such sedimentation processes leads to new fundamental insights into, for example, the formation of stars \cite{Caiazzo:2021}, the structure of earth layers in geology \cite{Dirks:2010}, evolutionary biology \cite{beaufort:2022}, archaeology \cite{Vernot:2021}, and climate science \cite{Turbet:2021}. It also benefits applications such as the characterisation of (macro)molecules in pharmaceutics \cite{Gabrielson:2011}, water purification \cite{Formentini:2013}, and revealing the equation of state of a system \cite{Piazza:1993}. In some cases, sedimentation leads to counter-intuitive effects, such as in the Brazil-nut effect, where under certain conditions, larger particles in a granular mixture move to the top, while smaller particles travel downwards upon shaking the system. Here, we experimentally demonstrate that a similar effect can occur in suspensions of charged colloids as an equilibrium phenomenon without needing external energy input. Furthermore, we theoretically obtained new insights into the Brazil-nut effect in the case of strongly interacting concentrated colloidal systems.

\begin{figure*}
\begin{center}
\includegraphics[width=\textwidth]{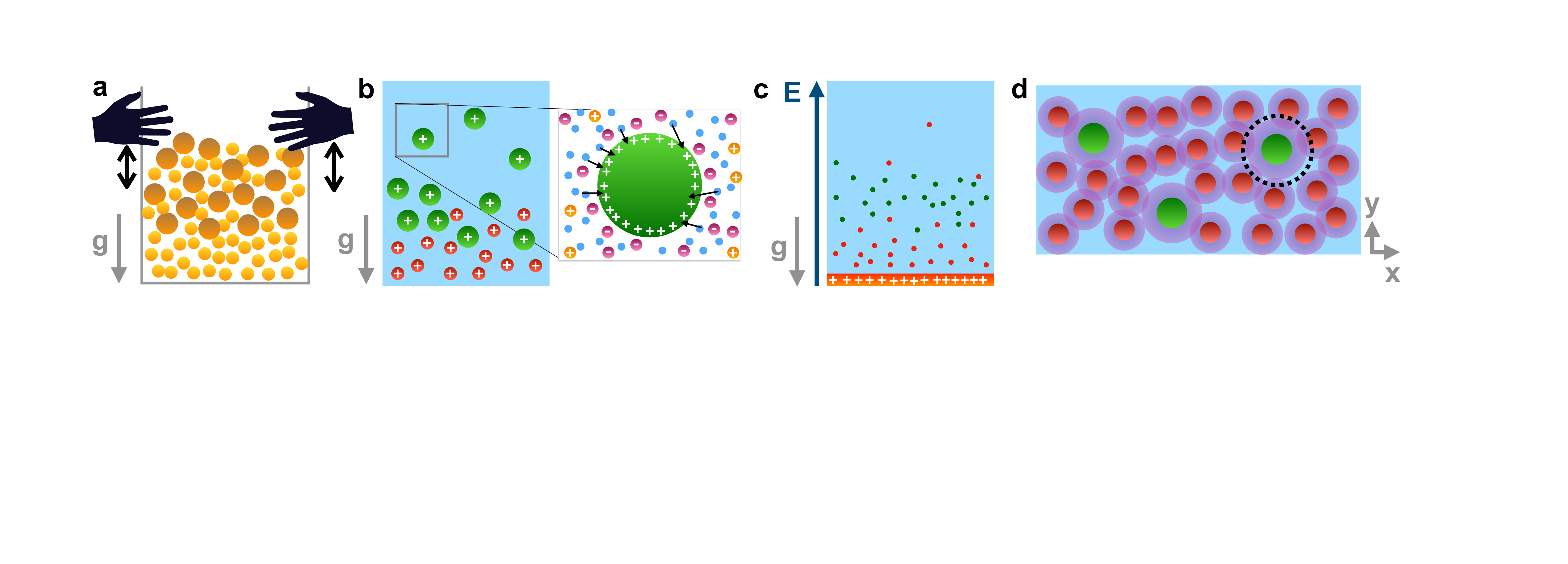} 
\end{center}
\caption{\label{fig:cartoon} {The Brazil-nut effect in granular and charged colloidal systems.} (a) In granular systems, particles can move against the gravitational force because of the input of external energy (e.g. via shaking). (b) No external driving is needed for charged colloids: the charged particles exhibit Brownian motion due to collisions with the surrounding solvent molecules and ions (inset). There are two different explanations for the occurrence of a Brazil-nut effect in such systems (c) At low salt concentrations, the entropic cost for counterions to establish local charge neutrality throughout the sample is too high, which causes a violation of local charge neutrality at the bottom of the container where the overall particle density is highest. This charge build-up self-consistently generates a macroscopic electric field ${\bf E}$ that pushes charged particles against the gravitational force. (d) In an effective picture of charged particles with an electric double layer (depicted in purple), a depletion layer is formed around the larger higher-charged particle at altitudes where there are more small than big particles. The particle plus depletion layer (dotted circle) has an effective density that is lower than the surrounding fluid consisting of solvent and small lower-charged particles, causing an upward buoyant force.}
\end{figure*}

The Brazil-nut effect was named after the large Brazil nuts that rise to the top of a mixture of nuts of different sizes when it is shaken \cite{williams76,rosato87}, see Fig. \ref{fig:cartoon}a. In colloidal systems, the Brazil-nut effect has been observed in systems consisting of active particles, which require continuous {external} energy input to maintain the activity \cite{Lowen:2019}. The colloidal phenomenon was named after the granular Brazil-nut effect only because a similar effect occurs: heavier particles float on top of the lighter ones. Density-functional theory \cite{esztermann04,zwanikken05,biesheuvel05,dijkstra06} and computer simulations \cite{esztermann04,dijkstra06,cuetos06} predict that the same effect can occur in charged colloidal systems, driven purely by electrostatics and Brownian motion (Fig.\ \ref{fig:cartoon}b), without the need for activity.  Surprisingly, despite the numerous predictions in theoretical and numerical studies, this effect has not yet been observed experimentally. {We stress that although the granular and charged colloidal Brazil-nut effects are analogous in the sense that they both lead to unusual density sorting, the mechanisms behind the effects are different.}

In charged colloids, two theoretical approaches have predicted the Brazil-nut effect. The first explanation uses the entropic lift effect that has been experimentally and theoretically studied for monodisperse suspensions  \cite{vanroij03,hynninen04, biesheuvel04, belloni05,torres07, biben94,rasa04,royall05,rasa05} and in mixtures of charged colloidal particles \cite{esztermann04,zwanikken05,biesheuvel05,dijkstra06,cuetos06}. At low ion concentration, it is entropically favourable for the system to locally violate charge neutrality in a {thin} layer at the bottom {and top} of the container to generate a macroscopic electric field that pushes charged colloidal particles to higher altitudes (Fig. \ref{fig:cartoon}c), defying gravity. Without this electric field, the high local density of colloidal particles at the bottom of the container would require {a highly inhomogeneous counterion concentration} to maintain local charge neutrality, which is entropically costly. The entropic cost to maintain local charge neutrality throughout the sample decreases at high ion concentrations. Therefore, the macroscopic electric field vanishes and the particles sediment according to a barometric distribution. 

A minimal density functional theory that captures the entropic lift effect, in which the components of the colloid-ion mixture are treated as massive colloidal particles and massless ions, found that the colloidal species separated into layers, such that colloids with the same mass-per-charge $m_i/Z_i$ were found at the same height \cite{zwanikken05}. Here $m_i$ is the buoyant mass, and $eZ_i$ is the charge of particles of species $i$ with $e$ the elementary charge. This is equivalent to the same value of $Z_iL_i$, with $L_i=m_ig/(k_\mathrm{B}T)$ the gravitational length, where $k_\mathrm{B}$ is the Boltzmann constant, $T$ temperature, and $g$ the gravitational acceleration. Particles with the lowest $m_i/Z_i$ are furthest from the surface onto which the particles sediment. From this ordering according to $m_i/Z_i$ it follows that the Brazil-nut effect occurs for $Z_\mathrm{L}/Z_{\mathrm{S}}\gtrsim m_\mathrm{L}/m_\mathrm{S}$, or $m_{\mathrm{L}}/Z_\mathrm{L} \lesssim m_\mathrm{S}/Z_\mathrm{S}$, i.e.\ the mass-per-charge for the heavier (large) colloids (L) is smaller than for the lighter (small) colloids (S). From this condition, it is clear why it is challenging to observe the Brazil-nut effect in charged colloids: it is non-trivial to introduce large charge ratios between the two particle species in the system and at the same time have low ion concentrations (i.e., large Debye screening lengths).

Later, the lift effect was confirmed in simulations that were either based on the primitive model \cite{hynninen04,torres07}, which includes colloids and ions as separate species, or used an approach in which colloids interact through an effective screened Coulomb potential \cite{torres07}, where the presence of the ions is only taken into account via the Debye screening length. The latter observation might come as a surprise at first because, in such an effective colloids-only description, the ions are integrated out, and they can, therefore, not redistribute such that a macroscopic electric field is generated. This apparent inconsistency can be understood by comparing the approximations in both approaches \cite{belloni05}. 

Within the alternative effective-potential approach, an intuitive explanation for the multi-component case was proposed in Ref. \cite{kruppa12} for the occurrence of a Brazil-nut effect valid for any system with long-range repulsions: at altitudes where more small low-charged particles than large high-charged particles are present, depletion layers are formed around the large particles because the repulsion between large and small particles is stronger than that between the small particles (Fig. \ref{fig:cartoon}d). The particle with its depletion layer consisting of solvent and ions is treated as an effective particle with lower mass density than the surrounding effective fluid composed of small particles, ions, and solvent. Therefore, due to buoyancy, the particles move to higher altitudes. At high ion concentrations, the depletion layer is reduced due to stronger electrostatic screening, which causes the Brazil nut effect to disappear. In addition, the depletion zone around the more repulsive particles results in an effective attraction of these particles to the hard bottom wall of the container, creating a layer of more repulsive particles on the container wall \red{\cite{Gon:1991,kruppa12}}. 

The effective-density picture agrees with experimental findings in Ref.\ \cite{gonzalezserrano11} and, more recently, Ref. \cite{Spruijt:2020} on sedimentation in charged colloidal mixtures. However, in both experimental works, no Brazil-nut effect was observed, presumably because the system parameters were not sufficiently extreme to be in the Brazil-nut regime. Interestingly, the description based on the macroscopic electric field already works on the level of point particles, excluding the possible formation of depletion layers.

In this article, we overcome the challenges of observing the colloidal Brazil-nut effect by using an experimental system of charged colloids in a low-polar solvent at low salt concentrations. Our findings support the prediction that for the Brazil-nut effect to occur, the mass-per-charge for the large colloids should be smaller than for the small colloids ($m_\text{L}/Z_\text{L}\lesssim m_\text{S}/Z_\text{S}$) \cite{esztermann04,zwanikken05,dijkstra06,cuetos06}. Furthermore, we quantitatively explain the experimental density profiles using a minimal dynamical density functional theory where the colloidal particles are modelled as constant-charge spheres. Finally, we discuss how theoretical predictions change when charge regulation --where the particle charge is determined by an equilibrium constant-- is added as an extra ingredient to the model, and we make predictions for future experiments.

\begin{figure}
\begin{center}
\includegraphics[width=0.43\textwidth]{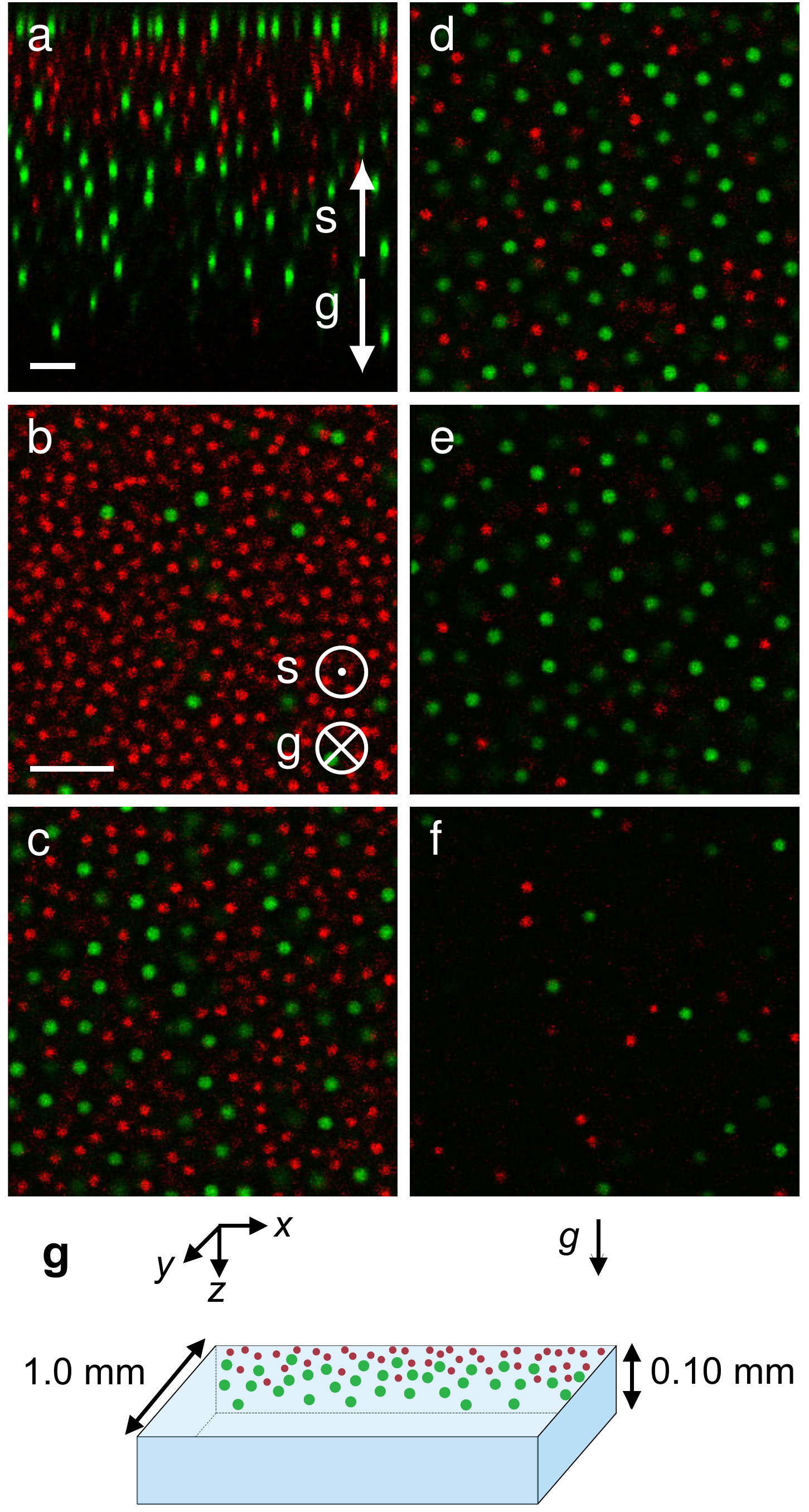} 
\end{center}
\caption{\label{BNfig:bn} {Colloidal Brazil-nut effect in a binary suspension of large (green) and small (red) particles} (system 2 in Table I of the SI; overall volume fraction $\bar{\eta}=0.02$). Gravity ($g$) points downwards, but the direction of sedimentation (s) is upwards, as the mass density of the PMMA particles is lower than that of the solvent CHB. (a)~$xz$~confocal image showing that the large (green) particles stay underneath the small (red) particles. (b)--(f) Sequence of $xy$ confocal images taken from top to bottom (spaced 12 $\mu$m apart; first one taken 8 $\mu$m from the top). Scale bars indicate 10 $\mu$m. (g) Schematic overview of the sample. $x$, $y$ and $z$ directions are indicated, as well as the direction of gravity ($g$). In reality, particles were present across the capillary's entire width ($y$), but for clarity, the particles are shown here only against the back wall ($xz$).}
\end{figure}

\section*{Results}
\subsection*{Experimental observation of the colloidal Brazil nut effect}
To arrive at the rather ``extreme" conditions for the Brazil-nut effect, particles are needed with large charge ratios compared to the ratio of their buoyant masses, at low ion concentrations (large Debye screening lengths). The latter is established by using a low-polar solvent (cyclohexyl bromide, CHB) as dispersing medium, which naturally contains a low {concentration} of ions \cite{yethiraj03}. We prepared four binary systems of micrometre-sized and sterically stabilised poly(methyl methacrylate) (PMMA) particles in CHB (see Materials and Methods) with different buoyant mass ratios $m_\text{L}/m_\text{S}=\sigma_\text{L}^3/\sigma_\text{S}^3$, where $\sigma_\text{L}$ and $\sigma_\text{S}$ are the diameters of the large and small particles, respectively, neglecting slight polydispersity. We also determined the charge (number) ratios $Z_\text{L}e/(Z_\text{S}e)=Z_\text{L}/Z_\text{S}$, with $Z_\text{L}e$ and $Z_\text{S}e$ the respective charges of the large and small particles. In choosing the particles for the binary mixtures, we exploited our observation that the covalent locking of the steric stabiliser (PHSA-PMMA) molecules to other PMMA chains forming the core of the particle (see Materials and Methods) has a profound effect on the particle charge \cite{Blaaderen:2015}. The phase behaviour and electrophoresis measurements indicate that locked particles carry a higher charge than the same particles in the unlocked state.

First, we test the theoretical prediction of the occurrence of the Brazil-nut effect, namely when $Z_\text{L}/Z_\text{S}\gtrsim m_\text{L}/m_\text{S}$. We define the Brazil-nut effect by the condition $h_\mathrm{L}>h_\mathrm{S}$, with $h_i$ the mean height of species $i$; see Materials and Methods. For cases where the ratio of buoyant masses is too large or when the charge ratio is too low, no Brazil-nut effect was found; see the SI Appendix for details.

\begin{figure*}
\begin{center}
\includegraphics[width=0.95\textwidth]{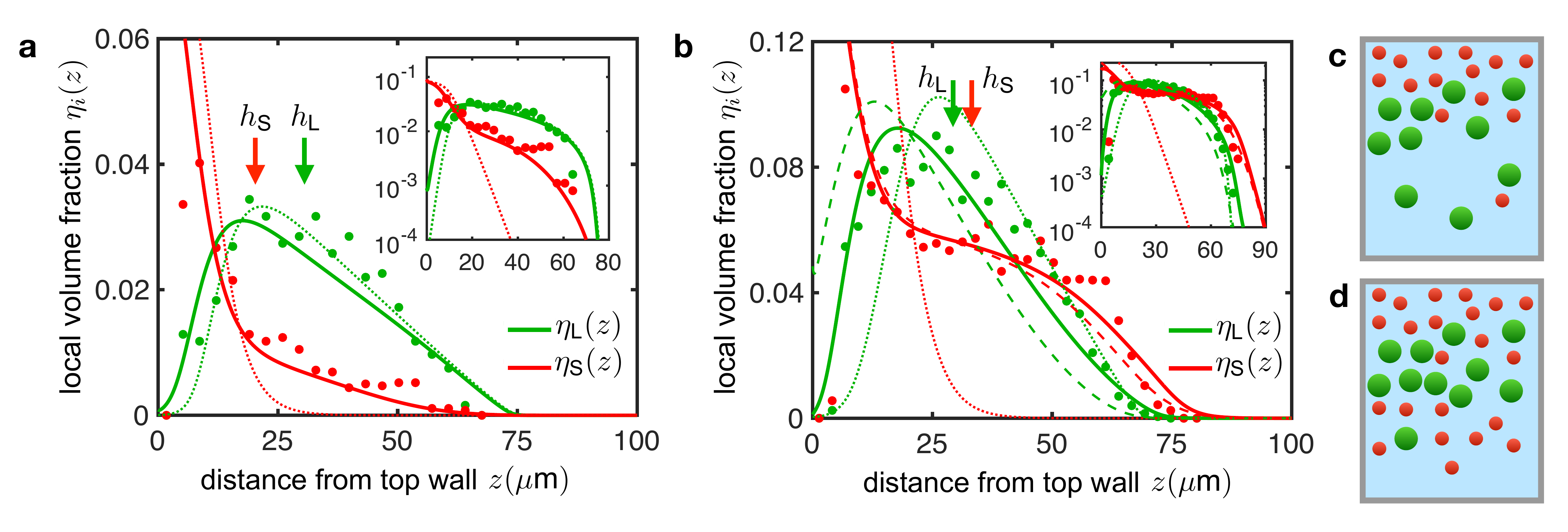} 
\end{center}
\caption{\label{BNfig:densprofs} {Density profiles for large (green) and small (red) particles in two systems} (corresponding to system 2 in Table I in the SI) with overall volume fractions (a) $\bar{\eta}=0.02$ and (b) $\bar{\eta}=0.07$. The data points are an average of 7 stacks. The arrows indicate the mean heights $h_\text{L}$ and $h_\text{S}$ of the large and small particles (see Materials and Methods, Eq.\ \eqref{BNeq:meanh}). The dotted lines are the best theoretical predictions using equilibrium theory, whereas the full and dashed lines are based on out-of-equilibrium theory. In (b), the difference between the full and dashed line is that for the full lines, an external wall potential is added to mimic the extra repulsion due to the adsorption of particles on the top wall. See Fig.\ \ref{BNfig:bn}g for an overview of the capillary, the wall onto which the particles sediment is located at $z=0$. Insets show the same data but on a log-linear scale. Used parameters of the theory are discussed in the main text. (c) Graphical depiction of the $\bar{\eta}=0.02$ and (d) the $\bar{\eta}=0.07$ sample.}
\end{figure*}

Next, we focus on a system where $Z_\text{L}/Z_\text{S}$ varies between 1.9 and 2.8 ($Z_\text{S}\approx240$), depending on the sample, and $m_\mathrm{L}/m_\mathrm{S}=2.0$, meaning that the condition for the Brazil-nut effect to occur is satisfied. $Z_\text{L}/Z_\text{S}$ was estimated by comparing one-component systems, which we expect to give a reasonable estimate for a 1:1 mixture, based on binary cell model calculations (see SI Appendix). Fig.\ \ref{BNfig:bn} shows confocal images that were taken for a sample (system 2) with overall volume fraction $\bar{\eta}=0.02$, which indeed exhibited the Brazil-nut effect.
 
Fig.\ \ref{BNfig:bn}a is an $xz$ cross-section of the capillary with the scale bars indicating 10 $\mu$m. The large green particles at the top of the image are adsorbed to the top wall of the capillary; the bottom wall is not visible in the image. Gravity points downwards; the particles, however, sediment upwards, as they have a lower mass density ($d_{\text{PMMA}}=$ 1.19 g cm$^{-3}$) than the solvent ($d_{\text{CHB}}=$ 1.336 g cm$^{-3}$). Because of the choice of the coordinate system (Fig. \ref{BNfig:bn}g), we can still use the intuition developed for systems with a positive buoyant mass (Fig. \ref{fig:cartoon}). We see that the large green particles ($L_\text{L}=0.71$~$\mu$m) are on average, further from the top wall than the small red particles ($L_\text{S}=1.39$~$\mu$m). Theory \cite{esztermann04,zwanikken05} predicts that charged particles are ordered according to their mass-per-charge $m_i/Z_i$, with particles with the lowest mass-per-charge furthest from the wall towards which the particles sediment (here: the top wall). This agrees with the buoyant masses and electrophoresis results (SI Appendix).

Figs.\ \ref{BNfig:bn}b--f are $xy$ confocal images taken parallel to the top wall of the capillary, spaced 12 $\mu$m apart, starting 8 $\mu$m from the top (Fig.\ \ref{BNfig:bn}b). Again, it is clear that the small red particles are closer to the top wall than the large green particles. Furthermore, the small particles have a smaller interparticle spacing than the large particles, consistent with a higher charge on the large particles. See also Supplementary Movie 1. Note, furthermore, the existence of depletion layers around the green particles (Fig. \ref{BNfig:bn}b) that were theoretically predicted \cite{kruppa12}.

\subsection*{Density profiles and comparison with theoretical model}

Fig.\ \ref{BNfig:densprofs} shows two experimentally determined density profiles for two samples of the system of Fig. \ref{BNfig:bn} (filled dots), at two different overall volume fractions, $\bar{\eta}=0.02$ (Fig.\ \ref{BNfig:densprofs}a) and $\bar{\eta}=0.07$ (Fig.\ \ref{BNfig:densprofs}b), with graphical depictions in Figs. \ref{BNfig:densprofs}c and d, respectively. The density profiles were calculated through $\eta_i(z)=\pi \sigma_i^3 \rho_i(z) / 6$, where $\eta_i(z)$ and $\rho_i(z)$ are the volume fraction and the number density as a function of the distance from the wall $z$. Images were taken 1-2 days after sample preparation. We excluded from the analysis the particles that were adsorbed to the wall (Fig. \ref{BNfig:bn}a), either via electrostatic interactions between the positively charged particles and negatively charged wall or via a boundary layering effect \cite{kruppa12}. 

For the $\bar{\eta}=0.02$ sample we found mean heights of $h_\text{L}=31$ $\mu$m and $h_\text{S}=20$ $\mu$m (Fig.\ \ref{BNfig:densprofs}a); for the $\bar{\eta}=0.07$ sample we found $h_\text{L}=30$ $\mu$m and $h_\text{S}=32$ $\mu$m (Fig.\ \ref{BNfig:densprofs}B). The gravitational lengths for the two species were $L_\text{L}=0.71$ $\mu$m and $L_\text{S}=1.39$ $\mu$m. For uncharged colloids in the dilute limit, a barometric profile is expected, with $h_i=L_i$ \cite{zwanikken05}; at higher densities, hard-core interactions would yield a more extended profile, but one would expect $h_i$ to be on the order of $L_i$ \cite{dijkstra06,cuetos06}. For our systems, the profiles were far more extended, with $h_i\gg L_i$, as would be expected for charged particles \cite{dijkstra06,cuetos06}.  

The $\bar{\eta}=0.02$ sample clearly displays the colloidal Brazil-nut effect with $h_\text{L}>h_\text{S}$. The $\bar{\eta}=0.07$ sample has $h_\text{L}\approx h_\text{S}$: this sample is approximately at the transition between Brazil-nut effect and no Brazil-nut effect. In Supplementary Movie 2, we show a stack of confocal $xy$ images of this sample for various values of $z$. 

To describe the experimental density profiles and explain the absence of a Brazil-nut effect for the sample with higher overall density, we employ a dynamical density functional theory for the colloid-ion mixture, where we include the ideal-gas entropy of colloidal particles and ions, mean-field electrostatics, and hard-core effects of the colloidal particles (see Materials and Methods). We assume that the ions react instantaneously to the colloid density field. The resulting Poisson-(Boltzmann)-Nernst-Planck equation gives for late times $t\rightarrow\infty$ (i.e. equilibrium) the same theory as discussed in Ref. \cite{esztermann04,zwanikken05}.  We chose this approach rather than an effective potential approach as in Ref. \cite{kruppa12} because our approach is simpler in the numerical calculation, allows easier implementation of charge regulation, and is known to perform better at low ion concentration \cite{torres07}. The trade-off of not using the effective potential approach is that depletion layers (see Fig. \ref{fig:cartoon}d dashed lines) are not taken into account, which is essential to describe particle adsorption on external walls by disruption of the otherwise spherical depletion layer \cite{kruppa12}. To overcome this specific shortcoming of the point-charge approach, we added external particle-wall potentials, $V_i(z)=\chi_i\exp(-\lambda z)$ with contact values $\chi_i$ ($i=\mathrm{L,S}$). Note that $\lambda^{-1}$ is a length scale associated with the size of the depletion layer, which depends on the ionic strength and particle charges that determine the inter-particle repulsion strengths between the same and different types of particles.

The experimental parameters with the highest uncertainty are the effective particle charges $Z_\mathrm{L}^0$ and $Z_\mathrm{S}^0$ of the homogeneous mixture and the reservoir Debye screening length $\kappa^{-1}$, and we systematically tune them to get the best match with the experimental data. Throughout, we set $\kappa^{-1}=1\ \mu\mathrm{m}$ to ensure that the dilute barometric regime within the sample cell is attained. This value is smaller than the experimental estimate based on conductivity measurements on the purified solvent CHB ($\kappa^{-1}=6\ \mu\mathrm{m}$). The Debye screening length being smaller than the experimental value can be rationalised from the increase in ionic strength in our systems due to a slight decomposition of CHB (see Ref. \cite{Blaaderen:2015}). To fix the particle charges, we used the data from electrophoresis measurements for one-component systems and estimated the particle charges in a 1:1 mixture (see SI Appendix) using a binary cell model \cite{Everts:2016}. We fixed the ratio to be $Z_\mathrm{L}^0/Z_\mathrm{S}^0=3$. The only remaining parameters that are changed to describe the experimental data are then the particle charges at a fixed charge ratio. Furthermore, due to the point-charge-like nature of the colloidal particles in the model, the charges should be treated as effective charges that are often lower than the experimental values \cite{Avni:2018}. The electrostatic potential in our calculation should therefore be viewed as a local Donnan potential. In a real system, the particle charges couple to the surface potentials rather than to the Donnan potential. A mapping between effective and experimental particle charges is possible using, for example, cell-model approaches (see SI Appendix). Other parameters are taken from the experiments (see Methods).

Surprisingly, the theory in equilibrium ($t\rightarrow\infty$) never quantitatively describes the density profiles, not even when the parameters in our theory are adjusted (Fig. \ref{BNfig:densprofs}, dotted lines); the theory predicts in equilibrium stronger particle segregation than was experimentally observed. However, our theory can give a quantitative agreement with the experimental density profiles (full lines, Fig. \ref{BNfig:densprofs}), provided that the density profiles are taken on an intermediate time $t\approx200\tau_\mathrm{D}$ for the low-density sample (Fig. \ref{BNfig:densprofs}a) and $t\approx80\tau_\mathrm{D}$ for the high-density sample (Fig. \ref{BNfig:densprofs}b). This observation suggests that equilibrium has not been reached in our experiments. Here, $\tau_\mathrm{D}=\sigma_\mathrm{L}^2/(4D)$ is the diffusion time with $D$ the diffusivity of the colloidal particles taken equal and density independent. For this system and our model, the equal diffusion constant approximation is not an unreasonable assumption, see SI Appendix. Given that hydrodynamic effects are also not explicitly accounted for we feel that trying to (only) correct the different diffusivities of the different species, without taking into account other effects such as concentration dependencies, will not yield additional insights. In this work, we are not after a quantitative mapping of the time scale.  

\begin{figure*}[t]
\centering
\includegraphics[width=\textwidth]{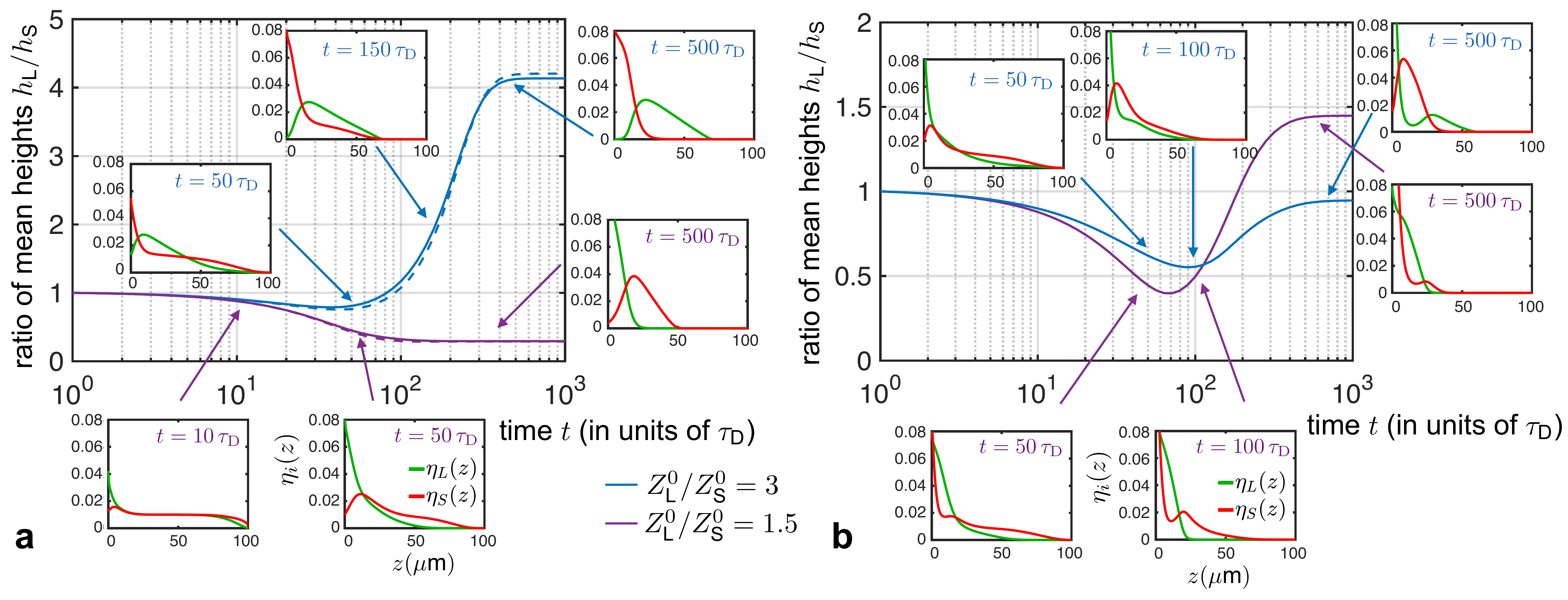}
\caption{{Sedimentation dynamics of a binary mixture of charged colloids} We quantify the mean height ratio $h_\mathrm{L}/h_\mathrm{S}$ and density profiles, as function of time. (a) We consider a charge ratio that results in a Brazil-nut effect in equilibrium ($Z_\mathrm{L}^0/Z_\mathrm{S}^0=3$, blue) for the constant-charge case (full lines) and the charge regulation case with only cation adsorption (dashed lines). We also consider a charge ratio with no Brazil-nut effect ($Z_\mathrm{L}^0/Z_\mathrm{S}^0=1.5$, purple). The insets show snapshots of the volume fraction profiles on times indicated by the arrows. (b) Same initial bulk charge ratios $Z_\mathrm{L}^0/Z_\mathrm{S}^0$ as in (a), but for particles where the large particles can adsorb anions (full blue line, $\alpha_\mathrm{L}=0.5$) suppressing the Brazil-nut effect, or when this only can happen for the small particle (full purple line, $\alpha_\mathrm{S}=0.6$), inducing the Brazil-nut effect. In both panels $\bar{\eta}_\mathrm{L}=\bar{\eta}_\mathrm{S}=0.01$ and $Z_\mathrm{S}^0=100$.}
\label{fig:time}
\end{figure*}

For matching the low-density experiment we used $Z_\mathrm{L}^0=300$ and $Z_\mathrm{S}^0=100$, whereas for the high-density experiment we used  $Z_\mathrm{L}^0=150$ and $Z_\mathrm{S}^0=50$. Both pairs of effective charges are expected to correspond well to the measured particle charges, see SI Appendix.  The lower bulk charges needed to describe the high-density experiment can be explained by a global discharging of particles upon increasing the overall density due to charge regulation (where the particle charge is described in terms of an equilibrium constant) \cite{Everts:2016}. For the high-density sample, we also needed the external particle-wall potentials to describe the experimental data because here we observed a strong particle absorption on the top wall. We take $\chi_\mathrm{L}/\chi_\mathrm{S}=Z_\mathrm{L}^0/Z_\mathrm{S}^0$ with $\chi_\mathrm{L}=50$ and $\lambda^{-1}=12.5\ \mu\mathrm{m}$. The theoretical prediction without such an external potential is shown by the dashed lines of Fig. \ref{BNfig:densprofs}b.

We obtain a lower bound for $\tau_\mathrm{D}$ using the single-particle diffusion coefficient calculated with the Stokes-Einstein relation. We find $\tau_\mathrm{D}=10$ s meaning that the density profiles in Fig. \ref{BNfig:densprofs} are reached within less than an hour, whereas on the experimental scale, the profiles were reached in a day.
More realistically, the single-particle diffusion coefficient should be replaced by a collective diffusion coefficient known to depend sensitively on particle density, the strength of electrostatic interactions, and hydrodynamic interactions \cite{Hess:1981, Dhont}. In particular, due to these effects, the collective diffusion coefficient has a lower value than the single-particle one, explaining at least qualitatively the gap between experimental and theoretical values. \blue{}

\subsection*{Metastable sedimentation profiles}
To understand why the high-density sample does not exhibit a Brazil-nut effect, we study in Fig. \ref{fig:time}a the generic sedimentation dynamics within our theoretical model for 
$Z_\mathrm{L}^0/Z_\mathrm{S}^0>m_\mathrm{L}/m_\mathrm{S}$ (blue line), and $Z_\mathrm{L}^0/Z_\mathrm{S}^0<m_\mathrm{L}/m_\mathrm{S}$ (purple line). The blue line is thus a system that is expected to give an equilibrium Brazil-nut effect, whereas the purple line does not. For both cases, we see a clear separation of time scales. When we view the  $Z_\mathrm{L}^0/Z_\mathrm{S}^0<m_\mathrm{L}/m_\mathrm{S}$ case, and probe the ratio of mean heights $h_\mathrm{L}/h_\mathrm{S}$ as a function of time, we see that there is never a Brazil-nut effect: $h_\mathrm{S}>h_\mathrm{L}$ for all times. One could hypothesise that the absence of a Brazil-nut effect in the high-density sample is because of too-low a ratio of the \emph{bulk} charges between large and small particles. However, when we look at the density profiles at various times for the purple curve in Fig. \ref{fig:time}a, none match the ones from the high-density experiment. In other words, the absence of a Brazil-nut effect for the high-density sample cannot be explained by a bulk charge ratio that is too small. 

For $Z_\mathrm{L}^0/Z_\mathrm{S}^0>m_\mathrm{L}/m_\mathrm{S}$ (blue line), there is initially no Brazil-nut effect ($h_\mathrm{L}<h_\mathrm{S}$) and the density profiles have the same shape as the one from the high-density experiment. However, for $t\gtrsim 100\tau_\mathrm{D}$ a Brazil-nut effect is found ($h_\mathrm{L}>h_\mathrm{S}$) and the shape of the density profiles changes to the one that strongly resembles the low-density experimental measurement. Then after around $t\approx 10^3\tau_\mathrm{D}$, equilibrium is reached with clear segregation of particles. 

It is tempting to conclude that we never found the equilibrium state in experiments, which would require longer waiting times. However, as the ionic strength keeps increasing due to the decomposition of the solvent, the equilibrium state at a constant Debye length of $\kappa^{-1}=1\ \mu\mathrm{m}, $ is experimentally unattainable. Furthermore, the system can freeze in a first-order phase transition at volume fractions where the colloids start to interact strongly or get arrested in an out-of-equilibrium glassy state. In both these limits, the collective diffusion coefficients become very small. Therefore, the high-density sample is expected to have a higher value of $\tau_\text{D}$ than the low-density sample. This gives a possible explanation for why there is no Brazil-nut effect in the high-density sample: after sufficiently long times, the sample is stuck in a metastable state with $h_\text{L}/h_\text{S}<1$, whereas the dynamics of the low-density sample is appreciably faster, such that a state of $h_\text{L}/h_\text{S}>1$ can be reached that is unattainable for the high-density sample, see Fig. \ref{fig:time}a blue line.  We note that the quantitative agreement between our highly simplified model and the experimental data could be fortuitous, for instance due to a cancellation of errors. However, considering the distinct sedimentation profile shapes generated as a function of time, we argue that our model captures quite reasonably the absence of a Brazil-nut effect in the high-density sample, which is the main reason of constructing the dynamical theory.

\subsection*{Effects of local charge regulation on sedimentation profiles}

We included the effects of charge regulation by adjusting the bulk charges; however, we neglected that charges can locally change due to an increase or decrease of the local densities compared to the bulk densities. One could argue that the absence of the Brazil-nut effect in Fig. \ref{BNfig:densprofs}b is because of such local discharging processes. To test this hypothesis, we extend our theory to include charge-regulation effects on the local level. First, we assume only charging of colloids due to cation adsorption, for which we specify the same bulk charges as the constant-charge case by fixing the value of the equilibrium constant appropriately (see Materials and Methods). Local discharging of particles at altitudes where the local density is larger changes the density profiles at low altitudes only quantitatively (see SI Appendix), following what is known in one-component systems \cite{biesheuvel04}. Qualitatively, no change was observed, as is seen from the $h_\mathrm{L}/h_\mathrm{S}$ curve as a function of time; in the dashed line of Fig. \ref{fig:time}a

When a particle can also adsorb anions, the discharging tendency is more significant, which causes more extreme local charge ratios. In our theory, we describe anion adsorption by the parameters $\alpha_i$ that represent the anion adsorption affinity for particle species $i=\mathrm{L,S}$, which is a free parameter even when we fix the bulk charges (see Materials and Methods). A high value of $\alpha_i$ means a large tendency for anions to adsorb; therefore the local charge ratio $Z_\mathrm{L}(z)/Z_\mathrm{S}(z)$ increases for $\alpha_\text{S}>0$ at low $z$, whereas it decreases for $\alpha_\text{L}>0$. Therefore, for all $t$ $\alpha_\text{L}>0$ suppresses a Brazil-nut effect that one expects to occur based on the bulk charges (Fig. \ref{fig:time}b, blue line), whereas at late times $\alpha_\text{S}>0$ can induce a Brazil-nut effect that would be absent based on bulk charges (Fig. \ref{fig:time}b, purple line). Besides inducing or suppressing a Brazil-nut effect, anion adsorption in these cases also drastically changes the shape of the density profiles, see the time snapshots in Fig. \ref{fig:time}b, to be compared with Fig. \ref{fig:time}a (where $\alpha_\mathrm{L,S}=0$). The change in the density profiles' qualitative features suggests that the colloidal particles' charging mechanism can be inferred from the shape of the density profiles in a sedimentation setup. Comparing the theoretical density profiles in Fig. \ref{fig:time} to the experimental density profiles (Fig. \ref{BNfig:densprofs}), we conclude that anion adsorption on the large particle does not play a prominent role in our experiments and that local charge regulation does not explain the absence of the Brazil-nut effect in Fig. \ref{BNfig:densprofs}b. 

\section*{Discussion}


Our combined experimental and theoretical approach demonstrates the realisation of a colloidal Brazil-nut effect in a non-active system driven purely by Brownian motion and electrostatics. Furthermore, the drastic change in shape when the charge regulation involves more than one ion species suggests that one can use sedimentation as a valuable tool to obtain microscopic information on the charging mechanisms of colloidal particles and possibly the value of the corresponding equilibrium constants. Finally, we also show that the occurrence of the Brazil-nut effect depends crucially on the overall density of the system, arguing that at higher density, slow dynamics prevented the system from reaching an equilibrium state that displays the Brazil-nut result. Our work, therefore, leads to new fundamental insights in partially glassy and crystalline systems \cite{Palberg:2009} and provides inspiration for future work. 

First, it would be interesting to experimentally and theoretically investigate the sedimentation dynamics of various charge-regulating particles with different charging mechanisms and see what kind of distinct time-resolved density profiles one can obtain as a function of salt concentration and overall density. {Concretely, it would be interesting to consider an experimental system in which either the large or the small particles can also adsorb anions and compare the sedimentation profiles to a system where only one species of ions can adsorb to the particles.} However, {the present work demonstrates} that the parameter space for observing the colloidal Brazil-nut effect may be limited in systems where charge regulation is essential, as is almost always the case in low-polar solvents. Namely, it was shown in Ref. \cite{Everts:2016} that if the charge density difference between the two colloidal species is too large, the charge on the higher-charged particle may induce oppositely charged patches on the other particle species, resulting in string formation and other phenomena that would frustrate the Brazil-nut effect \cite{Everts:2016}. Adding salt would provide better control of the screening length and make it possible to test the impact of the screening length on the density profiles in a binary system. Furthermore, it would be interesting to see how the density profiles' shape depends on how the sample has been prepared (initial conditions).

Second, from the theory viewpoint, it is surprising that our relatively simple mean-field dynamical density functional theory can quantitatively describe the experimental density profiles. In future work, it is interesting to include effects we have neglected here, such as hydrodynamic interactions, a renormalisation of collective diffusion coefficients due to direct interactions, and memory effects, such as caging. The latter is expected to play a prominent role: from Supplementary Movies 1 and 2, the particles are not freely diffusing. We could envisage that a more elaborate approach, for example, by using the many-body Smoluchowski equation \cite{Dhont}, can shed further light on why the metastable states from the experiments are observed and stable for such a long time and how the dynamics of a more complicated theory compares with our findings.  Here, confocal microscopy techniques can be a valuable asset, seeing that the full dynamics of the system can be probed using such techniques to extract the particles' self and collective diffusion coefficients.

\section*{Materials and Methods}
\subsection*{\label{BNsub:modelsystem}Model system}

We used poly(methyl methacrylate) spheres (PMMA; density $d_{\text{PMMA}}=$ \linebreak[4] 1.19~g~cm$^{-3}$; dielectric constant $\epsilon_\text{r}=2.6$; refractive index $n_\text{D}^{25}=1.492$ \cite{phdleunissen,elsesser10}, synthesised by dispersion polymerisation and sterically stabilised by a so-called comb-graft steric stabilising layer formed by poly(12-hydroxystearic acid) (PHSA) grafted onto a backbone of PMMA (PHSA-$g$-PMMA) \cite{bosma02,elsesser10}. We used four types of particles of different average diameters and labelled them with either the red fluorescent dye rhodamine isothiocyanate (RITC) or the green fluorescent dye 7-nitrobenzo-2-oxa-1,3-diazol (NBD). The two smaller particles had average diameters of 1.30 and 1.58 $\mu$m, polydispersities of 4.0\% and 3.5\%, respectively, and were labelled with RITC, the two larger particles had average diameters of 1.98 and 2.87 $\mu$m, polydispersities of 3.5\% and 2.4\%, respectively, and were labelled with NBD. The average diameters and polydispersities were determined by static light scattering (SLS) for the two smaller particles and by scanning electron microscopy (SEM) for the two larger particles. To determine the average diameter and size polydispersity from the SEM images, we measured $\sim100$ particles from each batch using the program iTEM (Olympus Soft Imaging Solutions GmbH). We note that the SLS diameter is generally a few per cent larger than the SEM diameter due to the swelling of the particles as a result of solvent uptake. Some batches of particles underwent a so-called `locking' procedure \cite{antl86}, in which the PMMA backbone of the steric stabiliser became covalently bonded to the particle surface; in this article, these particles are referred to as `locked'. In the case of `unlocked' particles, the PHSA-$g$-PMMA stabiliser is adsorbed to the particle surface but not covalently bonded. For more details on the characterisation of the PHSA-$g$-PMMA stabiliser and the chemistry involved in the locking step, which involves an additional heating step, see Ref.\ \cite{elsesser10}.

The particles were suspended in cyclohexyl bromide (CHB; Sigma-Aldrich; density $d_{\text{CHB}}=1.336$ g cm$^{-3}\ $  \cite{phdleunissen,yethiraj03}; dielectric constant $\epsilon_\text{r}=7.92$ \cite{heston50}; refractive index $n_\text{D}^{25}=1.4935$ \cite{heston50,phdleunissen}), which nearly matched the refractive index of the PMMA particles ($n_\text{D}^{25}=1.492$). 
This solvent is known to slowly and slightly decompose in time, a process which generates H$^+$ and Br$^-$ ions \cite{royall06}.
To reduce the ionic strength, we cleaned the solvent before use \cite{phdvissers,elsesser10} by bringing it into contact first with activated alumina (Al$_{2}$O$_{3}$; 58 \AA, $\sim$ 150 mesh, Sigma-Aldrich) and then with molecular sieves (4 \AA, 10--18 mesh, Acros Organics). 
The conductivity of CHB after the cleaning steps was on the order of $10$~pS cm$^{-1}$ (Scientifica 627 conductivity meter). 


\subsection*{\label{BNsub:electrophoresis}Electrophoresis measurements}

We measured the particles' electrophoretic mobility $\mu$ at a volume fraction $\eta\approx0.01$--0.02. These measurements are described in Ref. \cite{Blaaderen:2015}.
We used the theoretical work by Carrique et al.\ \cite{carrique02} (see also Ref.\ \cite{vissers11}) to calculate the electrostatic surface potential $\psi_0$ and charge number $Z$ from the mobility. In this work, a Kuwabara cell model is used to calculate $\psi_0$ and $Z$ from the measured mobility $\mu$ for any given screening length $\kappa^{-1}$ and volume fraction $\eta$ by numerically solving the full Poisson-Boltzmann equation. An estimate for $\kappa^{-1}$ was obtained from the measured conductivity of the solvent CHB (see below). Details can be found in Ref. \cite{Blaaderen:2015}.

\subsection*{\label{BNsub:dataacquisition}Sample preparation and confocal microscopy}

We prepared suspensions with an overall volume fraction of $\bar{\eta}=0.02$ or 0.07 and containing only one type of PMMA particle; note that upon sedimentation, the local volume fraction $\eta$ will be different from the overall volume fraction. We mixed equal volumes of the two suspensions and then transferred the resulting binary suspension to a borosilicate glass capillary (inner dimensions 5~cm~$\times$ 1.0~mm~$\times$ 0.10~mm ($x \times y \times z$); VitroCom) by dipping the capillary into the suspension. We also made samples by dipping the capillary into one of the suspensions and then into the second suspension, adding an approximately equal volume of the second suspension; mixing of the two suspensions took place during the filling step across a distance of a few micrometres and afterwards by particle diffusion. The suspension occupied approximately two-thirds of the capillary; the remaining part was left empty (containing only air). The capillary was mounted on a microscope glass slide, and both ends of the capillary were sealed with UV-curing optical adhesive (Norland no.\ 68), which also attached the capillary to the microscope glass slide. After curing, the sample was turned upside down (i.e.\ with the capillary below the microscope glass slide), which caused the two particle species to mix by forming swirls. In contrast, they sedimented to the opposite wall of the capillary (see, e.g.\ Refs.\ \cite{royall07,phdvermolen,milinkovic11}. The capillary was left to equilibrate for 1--2 days in this horizontal position with the $z$ axis parallel to gravity, as shown in Fig.\ \ref{BNfig:bn}G.

We used confocal microscopy (Nikon C1 or Leica SP2 confocal microscope) with a $63\times$ NA 1.4 oil immersion objective (Leica) in fluorescence mode with 543~nm (RITC) and 488~nm (NBD) excitation and sequential scanning mode, to obtain three-dimensional stacks of images (typical stack: 128 $\times$ 64 $\times$ 300 pixels ($x \times y \times z$); pixel size 0.20-0.24 $\mu$m;\linebreak[4] $\sim$ 3-6 frames per second). Fig.\ \ref{BNfig:bn}G is a schematic overview of the capillary with $x$, $y$ and $z$ directions and the direction of gravity ($g$) indicated.
Stacks were taken at several positions along the length ($x$ direction) of the capillary.

\subsection*{\label{BNsub:analysis}Data analysis}

We obtained the positions of the particles using an algorithm as described in, e.g. Refs. \cite{vanblaaderen95,dassanayake00}, which is an extension to 3D of the 2D method described in Ref.~\cite{crocker96}. 

From the particle coordinates, we calculated the number density profile $\rho_i(z)$ (number density as a function of the distance $z$ from the top wall of the capillary; see Fig.\ \ref{BNfig:bn}G) for each type of particle $i$. To improve statistics, we averaged the profiles from seven independent 3D stacks. We excluded particles that were adsorbed to the glass wall.  

To quantify the colloidal Brazil-nut effect, we can define, after Ref.\ \cite{esztermann04,zwanikken05}, for each density profile $\rho_i(z)$ for colloid $i$ a mean height
\begin{equation}
\label{BNeq:meanh}
h_i = \frac{ \displaystyle{\int_0^H} dz\, z \rho_i(z)}{ \displaystyle{\int_0^H} dz \, \rho_i(z) }\; , \;\; i= \text{L},\text{S}.
\end{equation}
The colloidal Brazil-nut effect is defined as $h_\text{L}>h_\text{S}$, with $h_\text{L}$ and $h_\text{S}$ the respective mean heights of the large and small colloids. 

\subsection*{\label{BNsubsub:scrl}Debye screening length}

The conductivity of CHB after the cleaning steps was on the order of $10$~pS $\text{cm}^{-1}$, which corresponds to an ionic strength of $c_\text{s}=2.4 \times 10^{-10}$ mol L$^{-1}$ (the ionic strength reduces for this case of a monovalent salt to the salt concentration $c_\text{s}$; note that the total ion concentration is $2c_\text{s}$).
We used Walden's rule to obtain an estimate for the ionic molar conductances for H$^+$ and Br$^-$ in CHB, which were needed to estimate the ionic strength from the measured conductivity. A Debye screening length $\kappa^{-1}\approx6$ $\mu$m follows. This calculation is described in more detail in Ref. \cite{Blaaderen:2015}. 

It is possible that $\kappa^{-1}$ was {smaller} in the samples for which we observed the Brazil-nut effect. This might be due to an increase of the ionic strength in time due to a slight decomposition of CHB.

\subsection*{\label{BNsubsub:electroph}Electrostatic surface potential and charge number}

From the measured electrophoretic mobilities, we obtained an estimate for the electrostatic surface potential $\psi_0$ and charge number per particle $Z$ (see Ref. \cite{Blaaderen:2015} for details on the procedure); the results are given in Table I of the SI Appendix. In Ref. \cite{Blaaderen:2015}, we found that the surface potential for locked particles was higher than for unlocked particles and that the charge increased quadratically with the particle diameter. In Ref.\ \cite{vissers11}, it was found for locked PMMA particles (diameter $\approx$ 1 $\mu$m) in a mixture of CHB and {\it cis}-decalin that at higher volume fractions ($\eta\gtrsim0.04$; depending on the system) the charge decreased significantly (for $\eta\approx0.13$  the charge was a factor of 1.5--2.0 lower than for $\eta\approx0.02$). The charge was approximately constant at low volume fractions ($\eta\lesssim0.04$; depending on the system).

To calculate the surface potential and the charge number from the measured electrophoretic mobility, we needed a value for the Debye screening length; we took $\kappa^{-1}\approx6$ $\mu$m, which was obtained from the conductivity of CHB directly before use, as described above. We did not measure the electrophoretic mobility for the unlocked particles U29 and U13. As the surface potential seemed approximately independent of the particle diameter \cite{Blaaderen:2015}, we assumed that the surface potentials of the particles U29 and U13 were similar to the surface potentials on the particles U20 and U16, namely $\beta e\psi_0\approx3.6$. From this, we calculated for particles U29 and U13 the charge numbers $Z_i=6.7 \times 10^2$ and $1.4 \times 10^2$, respectively.

The charge numbers of the particles in the samples for which we observed the Brazil-nut effect might have been different from the charge numbers obtained from electrophoresis due to charge regulation and differences in ionic strength, volume fraction and/or number ratio between the two species (see main text).

\subsection*{Dynamical density functional theory for constant-charge particles}
We use a theoretical model based on the dynamical density functional theory to describe the experiments. Neglecting hydrodynamic interactions and assuming the system is close to equilibrium, we find for $i=\mathrm{L,S}$,
\begin{equation}
\frac{\partial\rho_i({\bf r},t)}{\partial t}=\nabla\cdot \left\{ D_i\rho_i({\bf r},t)\nabla\left[\frac{\delta\beta\mathcal{F}}{\delta\rho_i({\bf r},t)}+\beta V_{i}({\bf r})\right]\right\}. \label{eq:DDFTc}
\end{equation}
In general, the diffusivities $D_\mathrm{L}$ and $D_\mathrm{S}$  depend on the local densities $\rho_\mathrm{L}({\bf r})$ and  $\rho_\mathrm{S}({\bf r})$, but for simplicity, we take it here as a constant with $D=D_\mathrm{L}=D_\mathrm{S}$.  The equal-constant approximation is not a limiting assumption for this system and model, see SI Appendix. As external potential, we take
\begin{equation}
\beta V_i(z)=\frac{z}{L_i}+\chi_i\exp(-\lambda z), \quad(i=\mathrm{L,S}) \label{eq:extpot}
\end{equation}
describing gravitational effects and that of an external wall potential due to particle adsorption onto the wall at $z=0$ \cite{Gonzalez:1992}.
 The intrinsic Helmholtz free energy functional of a binary colloidal dispersion with small (S) and large (L) charged colloidal particles in a 1:1 electrolyte with the solvent approximated as a continuum is
\begin{align}
\beta\mathcal{F}[\rho_\pm&,\rho_\mathrm{L},\rho_\mathrm{S}]=\sum_{\alpha=\pm}\int d{\bf r}\, \rho_\alpha({\bf r})\{\ln[\rho_\alpha({\bf r})\Lambda_\alpha^3]-1\} \nonumber
\\ &+\sum_{i=\mathrm{L,S}}\int d{\bf r}\, \rho_i({\bf r})\{\ln[\rho_i({\bf r})\mathcal{V}_i]-1\}\nonumber
\\ &+\frac{1}{2}\int d{\bf r}\, q({\bf r})\phi({\bf r})+\int d{\bf r}\, f_\text{HS}(\rho_\mathrm{L}({\bf r}),\rho_\mathrm{S}({\bf r})),
\label{eq:df}
\end{align}
with electrostatic potential $\phi({\bf r})/(\beta e)$ constrained by the Poisson equation
\begin{equation}
    \nabla^2\phi({\bf r})=-4\pi\ell_\mathrm{B}q({\bf r}),
    \label{eq:Poisson}
\end{equation}
with the total charge density is $eq({\bf r})=e[\rho_+({\bf r})-\rho_-({\bf r})+\sum_i Z_i\rho_i({\bf r})]$ and Bjerrum length $\ell_\mathrm{B}$. The integrations in Eq. \eqref{eq:df} are performed in a rectoid sample cell with height $H$ that is translationally invariant in the $xy$ plane. Finally, $\Lambda_\pm^3$ and $\mathcal{V}_\mathrm{L,S}$, are the (irrelevant) thermal volumes of ions and colloidal particles, respectively, and $f_\mathrm{HS}({\rho}_\mathrm{L},\rho_\mathrm{S})$ is the free energy density of a binary hard-sphere fluid, which in Eq. \eqref{eq:df} is taken in the local density approximation. Such contributions have been considered in sedimenting charged binary colloids in Ref. \cite{biesheuvel05}, although in a different parameter regime than the experiments of this work.

Explicitly writing out Eq. \eqref{eq:DDFTc} using Eqs. \eqref{eq:extpot} and \eqref{eq:df}, we find
\begin{align}
\frac{1}{D}\frac{\partial\rho_i(z,t)}{\partial t}=&\frac{\partial^2\rho_i(z,t)}{\partial z^2}+\frac{\partial}{\partial z}\Bigg\{\rho_i(z,t)\frac{\partial}{\partial z}\Bigg[Z_i^0\phi(z,t)\nonumber \\
&\beta\mu_i^\mathrm{HS}(\rho_\mathrm{L}(z,t),\rho_\mathrm{S}(z,t))+\beta V_i(z)\Bigg]\Bigg\}, \label{eq:PNP1}
\end{align}
where we defined $\mu_i^\text{HS}(\rho_\mathrm{L},\rho_\mathrm{S})=\partial_{\rho_i}f(\rho_\mathrm{L},\rho_\mathrm{S})$ for $i=\mathrm{L,S}$. We will take the Boublik-Mansoori-Carnahan-Starling-Leland equation of state \cite{Boublik:1970, Mansoori:1971}, with chemical potential,
\begin{align}
\beta\mu^\mathrm{HS}_i=&-\left(1+\frac{2\xi_2^3\sigma_i^3}{\eta^3}-\frac{3\xi_2^2\sigma_i^2}{\eta^2}\right)\ln(1-\eta) \nonumber \\
&+\frac{3\xi_2\sigma_i+3\xi_1\sigma_i^2+\xi_0\sigma_i^3}{1-\eta}+\frac{3\xi_2^2\sigma_i^2}{\eta(1-\eta)^2} \nonumber \\
&+\frac{3\xi_1\xi_2\sigma_i^3}{(1-\eta)^2}-\xi_2^3\sigma_i^3\frac{\eta^2-5\eta+2}{\eta^2(1-\eta)^3},
\end{align}
with $\sigma_i$ the particle diameters and total volume fraction $\eta=\eta_\mathrm{L}+\eta_\mathrm{S}$. Furthermore, we defined the quantity
\begin{equation}
\xi_k=\sum_{j=\mathrm{L,S}}\eta_j\sigma_j^{k-3}, \quad k=1,2,3,
\end{equation}
with volume fraction $\eta_i=(\pi/6)\sigma_i^3\rho_i$ ($i=\mathrm{L,S}$). The hard-sphere contribution is needed because, locally, the volume fraction can be larger than 0.1 in the parameter regime we will consider.

We assume that ions settle instantaneously for every time $t$ given the particle densities $\rho_\mathrm{L,S}({\bf r},t)$, and that they are treated grand canonically (ion particle number is not conserved) at chemical potential $\beta\mu_\pm=\ln(c_\mathrm{s}\Lambda_\pm^3)$, i.e. as if they are in contact with an ion reservoir at salt concentration $2c_\mathrm{s}$. This is reasonable because CHB acts as an ion reservoir that generates H$^+$ and Br$^-$ ions. Then the ions are Boltzmann distributed, $\rho_\pm({\bf r})=c_\mathrm{s}\exp[-\phi({\bf r})]$, combined with the Poisson equation Eq. \eqref{eq:Poisson}, results in the modified Poisson-Boltzmann equation
 \begin{equation}
\nabla^2\phi({\bf r},t)=\kappa^2\sinh[\phi({\bf r},t)]-4\pi\ell_\mathrm{B}\sum_{i=\mathrm{L,S}}Z_i^0\rho_i({\bf r},t), \label{eq:PNP2}
\end{equation}
with $\kappa^{-1}=(8\pi\ell_\mathrm{B}c_\mathrm{s})^{-1/2}$ the reservoir Debye screening length. Eqs. \eqref{eq:PNP1} and \eqref{eq:PNP2} take the form of modified Poisson-Boltzmann-Nernst-Planck equations and are solved with colloidal-particle blocking boundary conditions and imposed global charge neutrality. For the initial conditions, we use
 \begin{equation}
 \rho_i({\bf r},0)=\frac{\bar{\eta}_i}{(\pi/6)\sigma_i^3}, \quad (i=\mathrm{L,S}). \label{eq:t1}
 \end{equation}
From local charge neutrality at $t=0$, we find the initial condition for $\phi({\bf r},t)$ consistent with Eqs. \eqref{eq:t1},
\begin{equation}
\phi({\bf r},0)=\phi_\mathrm{D}=\mathrm{arsinh}\left\{\frac{\sum_iZ_i^0\bar{\eta}_i/[(\pi/6)\sigma_i^3]}{2c_s}\right\},
\label{eq:Donnan}
\end{equation}
with $\phi_\mathrm{D}/(\beta e)$ the Donnan potential of the homogeneous mixture.

In the model, we used experimental input values for the system size $H=100\ \mu\mathrm{m}$, Bjerrum length $\ell_\mathrm{B}=7$ nm, gravitational lengths $L_\mathrm{L}=0.71\ \mu\mathrm{m}$ and $L_\mathrm{S}=1.39\ \mu\mathrm{m}$, particle diameters $\sigma_\mathrm{L}=1.98\ \mu\mathrm{m}$ and $\sigma_\mathrm{S}=1.58\ \mu\mathrm{m}$, and the overall volume fractions of the homogeneous mixture $\bar{\eta}_\mathrm{L}$ and $\bar{\eta}_\mathrm{S}$. For the low-density experiments we used $(\bar{\eta}_\mathrm{L},\bar{\eta}_\mathrm{S})=(0.0125,0.01)$ and for the high-density experiment $(\bar{\eta}_\mathrm{L},\bar{\eta}_\mathrm{S})=(0.035,0.045)$, close to the experimental values. The values of the Debye screening length and bulk particle charges are discussed in the main text.

\subsection*{Charge regulation}
We extend the theory also to include charge regulation \cite{Ninham:1971, Avnipod:2019}. Therefore, we make the decomposition of the colloid charge density $\rho_i({\bf r})eZ_i({\bf r})$ in adsorbed positive ions $Z_{i,+}({\bf r})$ and adsorbed negative ions $Z_{i,-}({\bf r})$,
\begin{equation}
Z_i({\bf r})=Z_{i,+}({\bf r})-Z_{i,-}({\bf r}), \quad (i=\mathrm{L,S}).
\end{equation}
The total local particle charge is then determined by the adsorption of cations and anions on specific surface sites with total available sites for cations and anions, $M_{i,\pm}$. In the point-particle limit, the additional charge-regulation free energy that is added to Eq. \eqref{eq:df} is
\begin{equation}
\beta\mathcal{F}_\mathrm{CR}[\rho_\mathrm{L},\rho_\mathrm{S},Z_{\pm}]=\sum_{i=\mathrm{L,S}}\int d{\bf r}\, \rho_i({\bf r})\beta g_i(Z_{i,\pm}({\bf r})),
\label{eq:freeCR}
\end{equation}
with ``surface" free energy $g_i$ for finite size colloidal particles \cite{Everts:2017} treated in the point particle limit for $i=\mathrm{L,S}$,
\begin{align}
\beta g_i(Z_{i,\pm}({\bf r}))&=\sum_{\alpha=\pm}M_{i,\alpha}\Bigg\{\vartheta_{i,\alpha}({\bf r})\left[\ln\vartheta_{i,\alpha}({\bf r})+\ln\left(K_{i,\alpha}\Lambda_{\alpha}^3\right)\right] \nonumber\\
&+[1-\vartheta_{i,\alpha}({\bf r})]\ln[1-\vartheta_{i,\alpha}({\bf r})]\Bigg\},
\end{align}
describing a binary mixture of occupied and non-occupied sites with single-ion adsorption free energy $-k_\mathrm{B}T\ln(K_{i,\pm}\Lambda_\pm^3)$ and surface coverages $\vartheta_{i,\pm}$ for $i=\mathrm{L,S}$. 
Similar models for charge regulation for mobile charged colloidal particles were considered in the context of effective screening constants and their density profiles near a charged wall \cite{Markovich:2017}, and the effects on the disjoining pressure between two charged walls \cite{Avni:2018}.

Using that $Z_{i,\alpha}({\bf r})=M_{i,\alpha}\vartheta_{i,\alpha}({\bf r})$, we find from using the Euler-Lagrange equation
\begin{align}
\frac{\delta}{\delta\vartheta_{i,\alpha}({\bf r})}&\Bigg[\mathcal{F+F}_\mathrm{CR} \nonumber \\-\sum_{j=\mathrm{L,S}}&\int d{\bf r}\, \rho_j({\bf r})\sum_{\lambda=\pm}\mu_\lambda M_{j,\lambda}\vartheta_{j,\lambda}({\bf r})\Bigg]=0,
\end{align}
the Langmuir form which implies that ions follow equilibrium distributions for every fixed configuration of colloidal particles at any time $t$ for $i=\mathrm{L,S}$,
\begin{equation}
Z_{i,\alpha}({\bf r},t)=\frac{M_{i,\alpha}}{1+K_{i,\alpha}/\rho_\alpha({\bf r},t)}\approx\frac{M_{i,\alpha}\rho_\alpha({\bf r},t)}{K_{i,\alpha}}, 
\end{equation}
where in the last step, we made the realistic assumption that $Z_{i,\alpha}/M_{i,\alpha}\ll 1$, or equivalently for $i=\mathrm{L,S}$,
\begin{equation}
Z_i({\bf r},t)=\gamma_{i}c_s\Big\{\exp[-\phi({\bf r},t)]-\alpha_i\exp[\phi({\bf r},t)]\Big\}, 
\label{eq:CRZ}
\end{equation}
with charge regulation input parameters,
\begin{equation}
\gamma_i=\frac{M_{i,+}}{K_{i,+}}, \quad \alpha_i=\frac{M_{i,-}K_{i,+}}{M_{i,+}K_{i,-}}, \quad (i=\mathrm{L,S}).
\end{equation}
Note that $\gamma_i$ (units of inverse volume) determines the overall charge, whereas $\alpha_i$ characterises the tendency of negative ions to adsorb compared to the adsorption of positive ions. Also note that the assumption $Z_{i,\alpha}\ll M_{i,\alpha}$ reduces the number of input parameters from eight $(M_{i,+},M_{i,-},K_{i,+},K_{i,-})$ to four $(\gamma_i,\alpha_i)$. Eq. \eqref{eq:CRZ} is directly used in Eq. \eqref{eq:PNP2}. Furthermore, Eq. \eqref{eq:PNP1} is altered by the replacement
\begin{equation}
Z_i^0\phi(z,t)\rightarrow-c_s\gamma_i\Big\{\exp[-\phi(z,t)]+\alpha_i\exp[\phi(z,t)]\Big\}
\end{equation}
This should be contrasted with the implementation of charge regulation by Biesheuvel \cite{biesheuvel04} in the context of charge-regulating one-component colloids in a centrifugal field \cite{rasa04}, who used the constant-charge equations and then made the substitution $Z_i\phi({\bf r})\rightarrow Z_i({\bf r})\phi({\bf r})$ in Eq. \eqref{eq:PNP1}, which amounts to the assumption $\delta\mathcal{F}_\text{CR}/\delta\rho_i({\bf r})=0$ in our model. However, we will not adopt this assumption to be consistent with our free energy. In a few test cases of our numerical calculations, we established that the qualitative behaviour does not change if we use the Biesheuvel procedure.

Moreover, we set
\begin{equation}
Z_i({\bf r},0)=Z_i^0, \quad (i=\mathrm{L,S}), \label{eq:t2}
\end{equation}
where $Z_\mathrm{L}^0$ and $Z_\mathrm{S}^0$ are particle charges of a homogeneous mixture with composition $\bar{\eta}_\mathrm{L}$ and  $\bar{\eta}_\mathrm{S}$. Given $Z_i^0$, and $\alpha_i$, also $\gamma_i$ is uniquely determined,
\begin{equation}
\gamma_i=\frac{Z_i^0}{c_s[\exp(-\phi_\mathrm{D})-\alpha_i\exp(\phi_\mathrm{D})]}.
\end{equation}
Note that equilibrium constants are always positive, hence $\gamma_i>0$. All parameters are fixed by experiment, although the parameters $\kappa^{-1}$, $Z_\mathrm{L}^0$, $Z_\mathrm{S}^0$ are not precisely known. The quantities $\alpha_\mathrm{L}$ and $\alpha_\mathrm{S}$ are free parameters, however, we assume that $Z_i^0>0$, such that we have the inequality,
\begin{equation}
0<\alpha_i <\exp(-2\phi_\mathrm{D}).
\label{eq:treshold}
\end{equation}
The dynamical equations are solved using COMSOL Multiphysics.

\section*{Author contributions}
M.N.v.d.L. performed the experiments and analysed experimental data under the supervision of \mbox{A.v.B.};  \mbox{J.C.E.} performed the theoretical calculations under the supervision of R.v.R. All authors contributed to writing and discussing the manuscript.

\section*{Competing interests}
The authors declare no competing interests.

\section*{Data availability}
All data and material are included in the
article and/or SI Appendix.

\section*{Acknowledgements}
We want to thank Johan Stiefelhagen for synthesising the RITC-labelled PMMA particles, his help with the locking procedure and valuable discussions on PMMA synthesis, Gul\c sen Heessels-G\" urbo\u ga for synthesising the NBD-labelled PMMA particles; Niels Boon and Robert Hołyst are thanked for interesting discussions. J. C. E. acknowledges financial support from the Polish National Agency for Academic Exchange (NAWA) under the Ulam programme Grant No. PPN/ULM/2019/1/00257. M.N.v.d.L. acknowledges financial support via a Toptalent grant from the Dutch Research Council (NWO) funded by the Dutch Ministry of Education, Culture and Science (OCW). This work is part of the D-ITP consortium, a program of NWO that is also funded by the Dutch Ministry of OCW.
 \makeatletter
\renewcommand\@biblabel[1]{#1.}
\makeatother
\bibliography{literature1} 

\newpage
\newpage

\end{document}